\documentclass[12pt]{article}

\usepackage{amsmath,amsthm,amsfonts}
\usepackage{moreverb,url}
\usepackage{subfigure}
\usepackage{tikz}
\usetikzlibrary{decorations.pathmorphing,patterns}

\usepackage[colorlinks,bookmarksopen,bookmarksnumbered,citecolor=red,urlcolor=red]{hyperref}

\newcommand\BibTeX{{\rmfamily B\kern-.05em \textsc{i\kern-.025em b}\kern-.08em
T\kern-.1667em\lower.7ex\hbox{E}\kern-.125emX}}

\begin{document}


\title{Closed stable orbits in a strongly coupled resonant Wilberforce pendulum}

\author{Misael Avenda\~no-Camacho, Alejandra Torres-Manotas\\ and Jos\'e A Vallejo}

\newcommand{\Addresses}{{
  \bigskip
  \footnotesize

  M.~, Avenda\~no, \textsc{Departamento de Matem\'aticas, Universidad de Sonora (M\'exico),
    Hermosillo, Son. 83000}\par\nopagebreak
  \textit{E-mail address}, M.~, Avenda\~no: \texttt{misaelave@mat.uson.mx}

  \medskip

  A.~Torres, \textsc{Facultad de Ciencias,
    Universidad Aut\'onoma de San Luis Potos\'i (M\'exico), San Luis Potos\'i, SLP 78295}\par\nopagebreak
  \textit{E-mail address}, A.~Torres: \texttt{alejatorresm@gmail.com}

  \medskip

  J. A.~Vallejo (Corresponding author), \textsc{Facultad de Ciencias,
    Universidad Aut\'onoma de San Luis Potos\'i (M\'exico), San Luis Potos\'i, SLP 78295}\par\nopagebreak
  \textit{E-mail address}, J.A.~Vallejo: \texttt{jvallejo@fc.uaslp.mx}

}}

\maketitle

%
%

\begin{abstract}
We prove the existence of closed stable orbits in a strongly coupled Wilberforce
pendulum, for the case of a $1:2$ resonance, by using techniques of geometric singular 
symplectic reduction combined with the more classical averaging method of Moser.
\end{abstract}


\section{Introduction}

The Wilberforce pendulum is a physical device composed of a long suspended spring
with a mass attached at the lower end, which is free to rotate around the vertical
axis, twisting the spring through a non-linear coupling.

\begin{center}
\begin{tikzpicture}
\node[] (a) at (0,-0.1) {};
\node[] (b) at (2,2) {};
\node[circle,color=gray,fill=gray,inner sep=4] (c) at (0,0,-1) {};
\draw[decoration={aspect=0.3, segment length=3mm, amplitude=3mm,coil},decorate] (0,5) -- (a); 
\draw[decoration={aspect=0.3, segment length=1.5mm, amplitude=3mm,coil},decorate] (2,5) -- (b); 
\fill [pattern = north east lines] (-1,5) rectangle (3,5.2);
\draw[thick] (-1,5) -- (3,5);
\node[circle,color=gray,fill=gray,inner sep=4.1] (c) at (0,0,1) {};
\draw[ultra thick] (0,0,0.6) -- (0,0,-1);
\draw[ultra thick] (-1.1,0,-5.5) -- (0.9,0,-5.5);
\node[circle,color=gray,fill=gray,inner sep=4] (c) at (-1.1,0,-5.5) {};
\node[circle,color=gray,fill=gray,inner sep=4] (c) at (0.9,0,-5.5) {};
\end{tikzpicture}
\end{center}

It can be found in any Physics laboratory, where it is used to demonstrate the 
periodic motion arising from transference of energy between the two main modes
of oscillation. If the spring is initially stretched, with a certain initial
torsion, and then released from rest, the motion will start being dominated by
an `up and down' swinging, which gradually converts itself into a purely rotational 
oscillation of the hanging mass.

This striking motion is crucially related to the non-linear coupling between
both oscillating modes. In the usual setting, that coupling is as weakest as possible,
being given by a quadratic term in the generalized coordinates. To be more specific,
consider the spring to me massless, with elastic and torsional constants being
$\kappa$ and $\rho$, respectively. Let the moment of inertia of the hanging mass $m$
be $I$. Finally, denote by $x$ the elongation of the spring and by $y$ the torsion
angle. The complete Lagrangian in the case of a quadratic coupling is then
\begin{equation}\label{eq1}
L=\frac{1}{2}m\dot{x}^2+\frac{1}{2}I\dot{y}^2-
\left(\frac{1}{2}\kappa x^2 +\frac{1}{2}\rho y^2 +\varepsilon xy\right)\,.
\end{equation}

This case is well known, as general references we can cite \cite{Ko90,BM91,PZB09}.
In this work, we will be interested in the case of a \emph{stronger} coupling, namely,
one given by a quartic non-linear term. The Lagrangian \eqref{eq1} has then to be
modified to read
$$
L=\frac{1}{2}m\dot{x}^2+\frac{1}{2}I\dot{y}^2-
\left(\frac{1}{2}\kappa x^2 +\frac{1}{2}\rho y^2 +\varepsilon x^2y^2\right)\,,
$$
with its corresponding Hamiltonian
\begin{equation}\label{eq2}
H=\frac{1}{2m}p^2_x+\frac{1}{2I}p^2_y+
\frac{1}{2}\kappa x^2 +\frac{1}{2}\rho y^2 +\varepsilon x^2y^2\,,
\end{equation}
and our goal is to study the existence of closed stable motions.

As stated above, the problem falls within the reach of perturbation theory. On
physical grounds, it was to be expected that a mild non-linear coupling would lead to 
periodic motions, `inherited' from the two independent oscillation modes that would
exist in the absence of coupling, but the situation is not so clear in the presence
of a strong non-linear coupling, as these interactions typically lead to a chaotic 
evolution, as shown in many textbooks such as \cite{Ott02,Str15}.

The standard tool for determining the onset of chaos in any dynamical system 
(in particular, perturbed ones) is the construction of a suitable Poincar\'e 
section in the phase space; we do this in
the next section, showing the progressive destruction of the integrable tori
associated with increasing values of the perturbation parameter. 
In the case in which the non-perturbed system is integrable, the 
Kolgomorov-Arnold-Moser (KAM) theorem guarantees the persistence of some of these 
tori, thus proving the existence of stable periodic orbits winding around some of
the corresponding orbits of the unperturbed system \emph{in the absence of 
resonances}. In this work, however, we will be interested in the 
$1:2$ resonance
(although the ideas presented will work for an arbitrary $m:n$ resonance), so the
KAM theorem will not be applicable.

Aside from the KAM theorem, there are other techniques in perturbation theory that
are well suited to the kind of problem at hand. Our path here will be to put first
the system into normal form by using the Lie-Deprit method (see \cite{Dep69}), 
and then to apply a
singular geometric reduction (as in \cite{Cus94,CB97}) to pass to a reduced phase 
space where closed stable orbits can be detected either by applying Moser's theorem 
on reduction (see \cite{Mos70,CKR83}) or by determining the fixed points on a suitable
Poincar\'e surface. This approach has been applied successfully to a qualitatively 
different system, the Pais-Uhlenbeck oscillator, in \cite{AVV17} so it can be 
considered to be of a general nature.

Transforming the original perturbed Hamiltonian \eqref{eq2} $H=H_0+\varepsilon H_1$,
where
$$
H_0=\frac{1}{2m}p^2_x+\frac{1}{2I}p^2_y+
\frac{1}{2}\kappa x^2 +\frac{1}{2}\rho y^2
$$
is the sum of two independent oscillators,
into another one in normal form $N=H_0+\sum^\infty_{j=1}\varepsilon^jN_j$, where
$\{H_0,N_j\}=0$ for each $j\in\mathbb{N}$ (with $\{\cdot,\cdot\}$ the canonical Poisson bracket on the algebra of smooth functions on the phase space, $\mathcal{C}^\infty(\mathbb{R}^4)$), requires solving a set of equations known (following V. I. Arnold) as the homological equations. To this end, it is convenient to make use of
the averaging method, based on two averaging operators, $\left\langle\cdot\right\rangle$ and $\mathcal{S}$, acting on tensor fields defined on the phase space
manifold (in particular, on smooth functions on phase space). This allows us to
obtain recursive formulas for computing the sub-Hamiltonians $N_j$, $j\in\mathbb{N}$.
Moser's theorem extracts information about the existence of closed stable orbits
from the critical points of the first-order normal perturbation $N_1$ on the reduced
phase space, but when there are degeneracies (as it turns out to be our case) one has
to go over $N_2$ (by considering $N=H_0+\varepsilon (N_1+ \varepsilon N_2)$
as the new perturbed Hamiltonian). We will use the explicit expressions for $N_1$ and
$N_2$ deduced in \cite{AVV13}, which have the advantage of not relying on the 
introduction on action-angle variables; on the contrary, they only depend on the 
averaging operators mentioned above, and the canonical Poisson bracket. Finally,
from the study of the critical points of $N_1$ and $N_2$ on the reduced phase space, 
we will be able to conclude the existence of closed stable orbits in the Wilberforce
pendulum for any fixed value of the perturbation parameter $\varepsilon >0$.

Let us remark that averaging techniques have been used to study a different problem, 
the existence of periodic orbits in a perturbed \emph{weakly} coupled Wilberforce
pendulum, see \cite{BLM16}. There, the authors parametrize the periodic solutions of 
that perturbed system by the simple zeros of an associated system of nonlinear 
equations.

\section{Numerical analysis}

To make explicit the characteristic frequencies of the system $w_1,w_2$, let us introduce them through $\kappa=mw^2_1$ and $\rho =Iw^2_2$. The Hamiltonian is
\begin{align}\label{eq3}
H=& H_0+\varepsilon H_1 \nonumber\\
=&
\frac{1}{2}\left(
p^2_x+mw^2_1x^2+p^2_y+Iw^2_2y^2
\right)
+\varepsilon x^2y^2\,.
\end{align}

The corresponding Hamilton equations of motion are given by the first-order system
	\begin{equation}\label{eq4}
	\begin{split}
	&\dot{x}  = p_x, \\
	&\dot{p}_x = -\left(m\omega_1^2 x+2\varepsilon xy^2\right), \\
	&\dot{y} = p_y,\\
	&\dot{p}_y = -\left(I\omega_2^2y+2\varepsilon y x^2\right).
	\end{split}
	\end{equation}
	
We solve this system numerically with the symplectic velocity Verlet method, see \cite{Hol07}. The numerical scheme in this case is
	\begin{equation*}
	\begin{split}
	&x_{i+1}=x_i+k(p_x){}_i+\dfrac{k^2}{2} F_1\left(x_i, y_i;\varepsilon\right),\\
	&(p_x){}_{i+1}=(p_x){}_i +\dfrac{k}{2}\left( F_1\left(x_{i+1},y_{i+1};\varepsilon\right)+ F_1\left(x_i,y_j;\varepsilon\right)\right),\\
	&y_{i+1}=y_i+k(p_y){}_i+\dfrac{k^2}{2} F_2\left(x_i, y_i;\varepsilon\right),\\
	&(p_y){}_{i+1}=(p_y){}_i +\dfrac{k}{2}\left( F_2\left(x_{i+1},y_{i+1};\varepsilon\right)+ F_2\left(x_i,y_j;\varepsilon\right)\right),
	\end{split}
	\end{equation*}
where $k>0$ is the step size and we have written $F_1(x, y;\varepsilon)=-m\omega_1^2 x-2\varepsilon xy^2$, $F_2(x, y;\varepsilon)=-I\omega_2^2y-2\varepsilon y x^2$. 

\begin{figure*}
	\centering
	\subfigure[$\varepsilon=0$]{\includegraphics[scale=0.33]{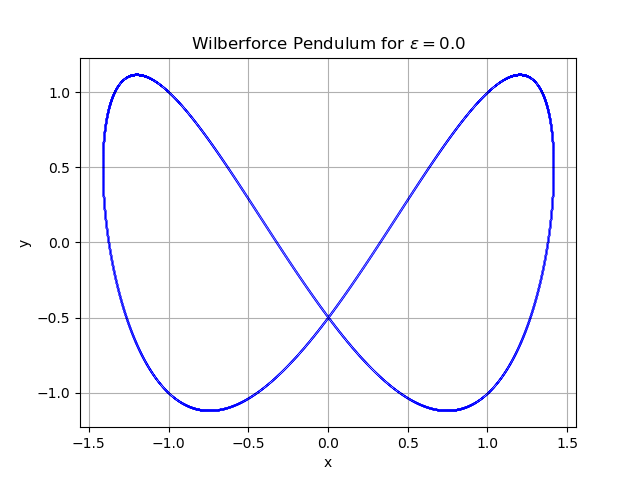}}
	\subfigure[$\varepsilon=0.4$]{\includegraphics[scale=0.33]{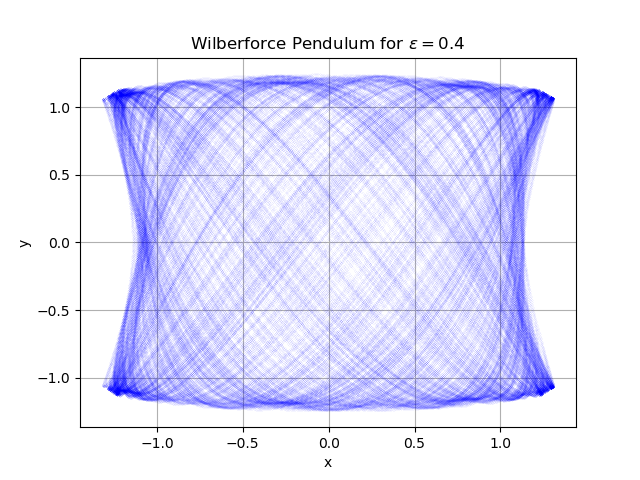}}
	\subfigure[$\varepsilon=0.5$]{\includegraphics[scale=0.33]{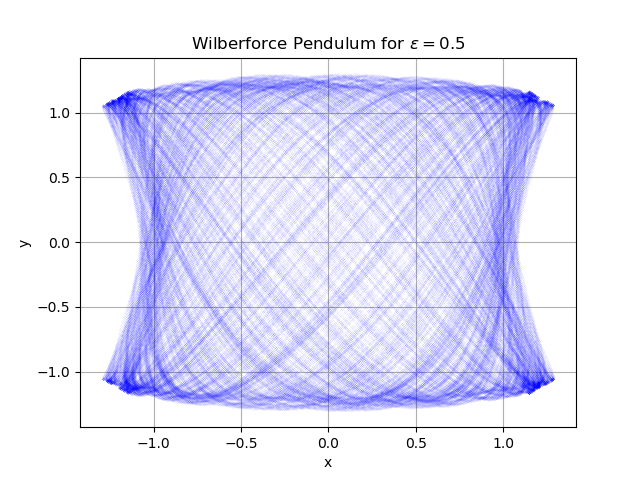}}
	\subfigure[$\varepsilon=0.6$]{\includegraphics[scale=0.33]{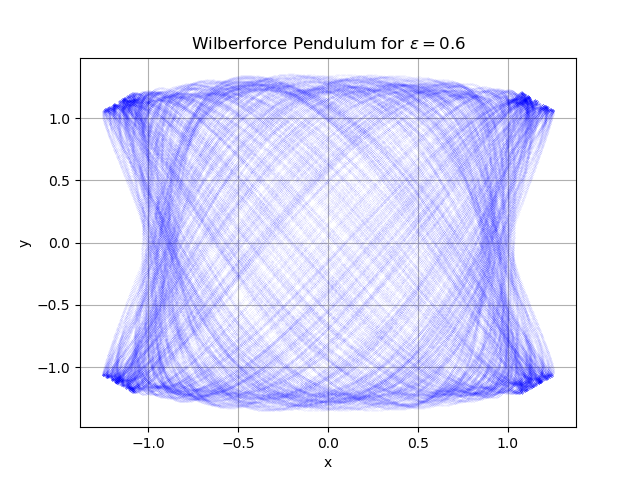}}
	\subfigure[$\varepsilon=0.7$]{\includegraphics[scale=0.33]{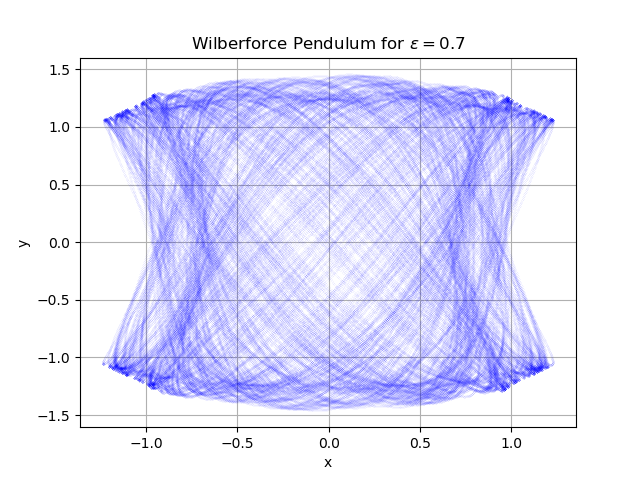}}
	\subfigure[$\varepsilon=0.8$]{\includegraphics[scale=0.33]{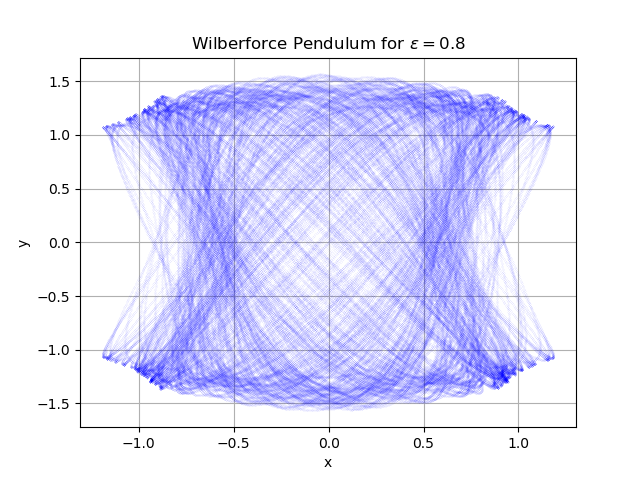}}	
	\caption{Strongly coupled Wilberforce pendulum dynamics for different values of 
	the parameter $\varepsilon$. The resonance 1:2 is shown.}
	\label{mov1}	
\end{figure*}

The resulting dynamics in the $xy$ plane is displayed in Figure \ref{mov1},
where we have taken as initial condition $(1,1,1,1)$, along with values 
$m=I=w_1=1$, $w_2=2$, which will be assumed in what follows. Notice that
the expected Lissajous figure appears when the coupling is switched off,
and the pattern becomes fuzzier around $\varepsilon =0{.}5$. 

The chaotic behavior is even more apparent when considering a Poincar\'e section.
We will take a surface $\sigma$ transversal to the flow of \eqref{eq4} constructed in 
the following way\footnote{Of course, this is just a choice. There are many 
possibilities for constructing a Poincar\'e surface, but the idea of the numerical
procedure is the same in all of them. Our choice is determined by reasons of graphic 
cleanliness.}: First, we fix an energy value $H=h$ in
\eqref{eq3}, and write $p_2$ in terms of $\left(p_1,q_1,q_2,h\right)$:

\begin{equation}\label{eq5}
p_2=\pm \sqrt{2\left( h-\varepsilon \left(q_1 q_2\right)^2\right)-\left(p_1^2+q_1^2+4q_2^2\right)}\,.
\end{equation}
Next, we restrict the solution to the level set $\Sigma_h:=\left\lbrace \left(p_1,q_1,q_2\right):H=h\right\rbrace$. The Poincaré section is then
$\sigma=\left\lbrace \left(p_1,q_1\right)\in \Sigma_h:q_2=0 \right\rbrace$.

\begin{figure*}
	\centering
	\subfigure[$\varepsilon=0.1$]{\includegraphics[scale=0.33]{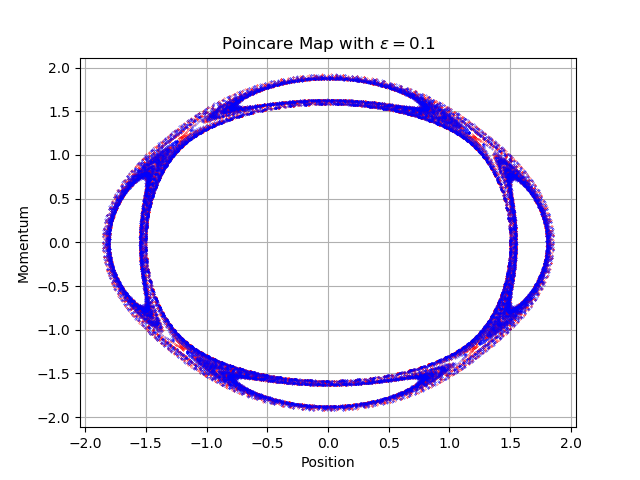}}
	\subfigure[$\varepsilon=0.2$]{\includegraphics[scale=0.33]{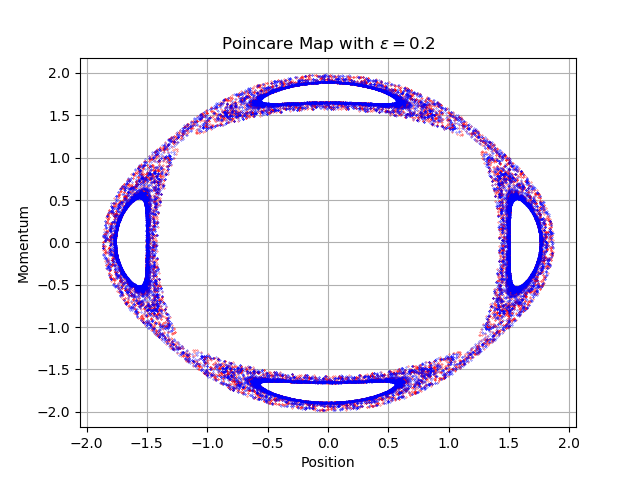}}
	\subfigure[$\varepsilon=0.4$]{\includegraphics[scale=0.33]{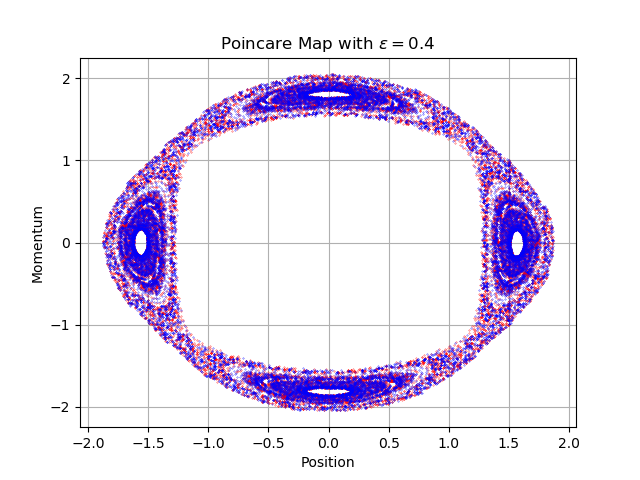}}
	\subfigure[$\varepsilon=0.5$]{\includegraphics[scale=0.33]{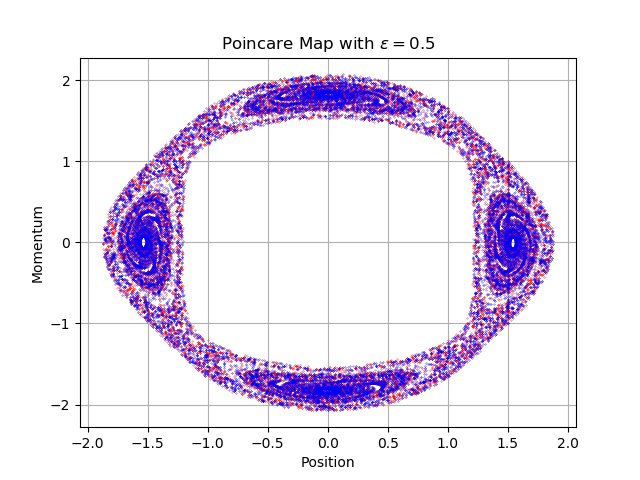}}
	\subfigure[$\varepsilon=0.8$]{\includegraphics[scale=0.33]{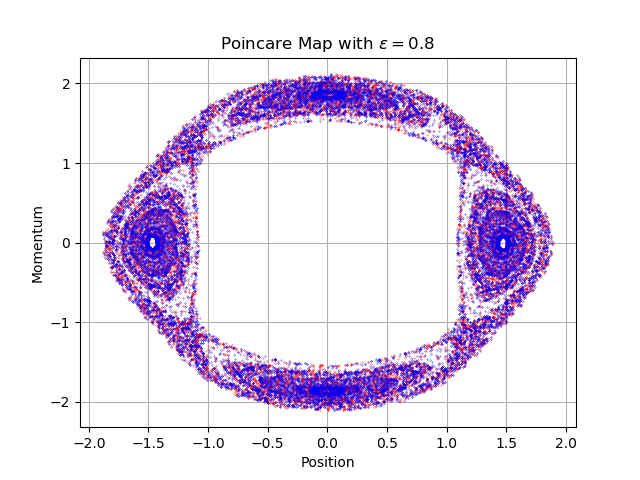}}
	\subfigure[$\varepsilon=0.9$]{\includegraphics[scale=0.33]{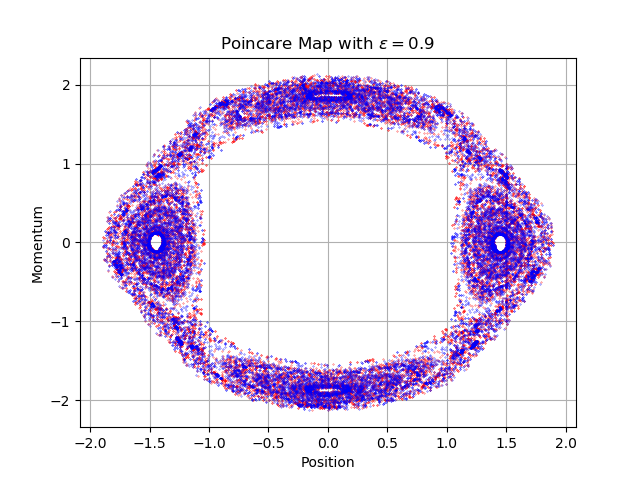}}
	\subfigure[$\varepsilon=1$]{\includegraphics[scale=0.33]{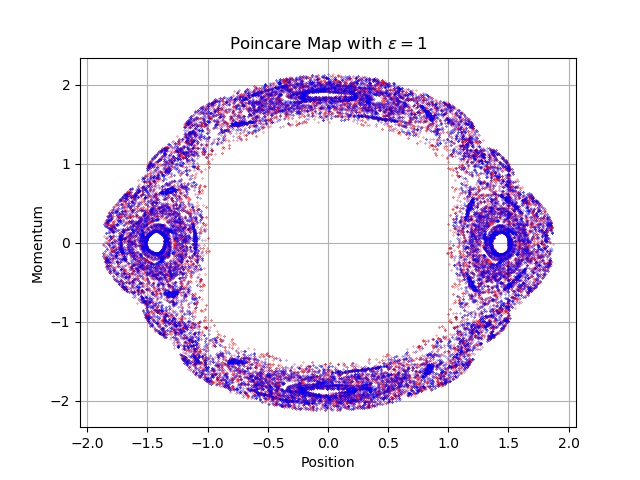}}
	\caption{Poincaré maps of the strongly coupled Wilberforce pendulum for different 
	values of $\varepsilon$. The resonance $1:2$ is shown.}
	\label{pcm1}	
\end{figure*}

Initial conditions are taken of the form $(j/100,1.5,p_2,0.01)$, 
where $j$ is a random number in $[-100,100]$ and $p_2$ is given in
\eqref{eq5}. In all cases, the value $h=3$ has been chosen.
Again, the evolution is computed with the velocity Verlet method, recording
the points that cross $\sigma$ by looking at sign changes. Notice that two set of solutions are obtained, one for each sign of \eqref{eq5}, which are superimposed to get the final Poincaré map for each value of the parameter $\varepsilon$. The results are shown in Figure \ref{pcm1}; as stated above, the destruction of the integrable tori is quite visible here. However, certain `islands of stability' survive (as in
the non-resonant case describes by KAM theorem), and in the next sections we prove
their existence analytically.

\section{Normal forms in perturbation theory}\label{sec-prelim}

Given a Poisson manifold $(M,\{\cdot ,\cdot\})$, consider a perturbed
Hamiltonian of the form $H=H_0+\varepsilon H_1$, where $H_0$ is supposed to be integrable.
Hamilton's equations for $H$ are  a coupled non-linear system of differential equations
whose solutions, in general, do not have a closed form. The Lie-Deprit approach to this problem substitutes the system of Hamiltonian equations by a simpler one, suitable to be studied by analytic tools, while providing some criterion to determine the degree of accuracy of the
approximation. The perturbed Hamiltonian $H$ is said to
admit a \emph{normal form} of order $n$ if there exist a near-identity canonical transformation
on phase space such that $H$ is transformed into
\begin{equation}\label{e1}
H=\sum^n_{i=0}\varepsilon^iN_i +R_H\,,
\end{equation}
where $N_0=H_0$ and
\begin{equation}\label{e2}
\{N_i,H_0\}=0,\mbox{ for all }1\leq i\leq n\,.
\end{equation}
The truncated function $N=\sum^n_{i=0}\varepsilon^iN_i$ is the normal form (of order $n$) of $H$.
This approach is based on the fact that whenever $\Vert H-N\Vert =\Vert R_H\Vert$ is small in a
suitable norm, the trajectories of $N$ provide us with good approximations to the true trajectories of $H$. In particular, closed orbits for $H$ can be detected through the 
existence of closed orbits for $N$.

The normal form is obtained from a family of canonical transformations
depending on the parameter $\varepsilon$, $x\mapsto y(x;\varepsilon)$ (where $x$ denotes
collectively the coordinates on $M$), such that $y(x;0)=x$. To assure that these transformations are canonical, they are derived from a generating function $S=S(\varepsilon)$:
\begin{equation}\label{e3}
\frac{\partial y_j}{\partial \varepsilon}=\{S,y_j\}=
\mathcal{L}_{X_S}y_j\,,\mbox{ for }j\in\{1,\ldots ,\dim M\}\,.
\end{equation}
with $X_S=\{S,\cdot\}$ the Hamiltonian vector field determined by $S$. 
Geometrically, $X_S$ is the `$\varepsilon-$flow generator', much in the same way as $H$ is the time-flow generator.

The Lie-Deprit method proceeds by developing $S$ in a formal series 
$S=\sum^n_{j=0}\varepsilon^jS_j$ and translating the condition of being a generating function 
for canonical transformations into a set of equations, one for each term $S_j$, having the structure
\begin{equation}\label{e4}
\mathcal{L}_{X_{H_0}}S_j=F_j-(j+1)N_{j+1}\,\mbox{ }j\geq 0\,,
\end{equation}
where the $F_j$ functions are determined by quantities already calculated in previous steps.
What is remarkable (see \cite{Dep69}) is that these equations (called the 
\emph{homological equations}) have a 
recursive structure (the Deprit's triangle) and they can be solved in terms of $H_0$ and the 
sub-Hamiltonians $N_j$. The usual method of solution is based on the introduction of action-angle
coordinates, thus having a local character and requiring a symplectic phase space. To avoid these 
issues here we follow \cite{AVV13}, where a global method of solution is presented in the case of
a system admitting a $U(1)-$action such that the Hamiltonian vector field $X_{H_0}$ has periodic 
flow, as is the case with the Wilberforce pendulum.

In a general setting, if we have a phase space which is a Poisson manifold $(M,\{\cdot,\cdot\})$,
given the Hamiltonian $H=H_0+\varepsilon H_1$ we can set up the homological equations
\eqref{e4}. Now, suppose that the vector field $X_{H_0}=\{H_0,\cdot\}$ is complete and has
periodic flow $\mathrm{Fl}^t_{X_{H_0}}$. 
The periodicity condition means that there exists a period
function $T:M\to\mathbb{R}$ such that 
$\mathrm{Fl}^t_{X_{H_0}}(p)=\mathrm{Fl}^{t+T(p)}_{X_{H_0}}(p)$. 
This flow induces a $U(1)-$action by putting 
$(t,p)\mapsto \mathrm{Fl}^{t/w(p)}_{X_{H_0}}(p)$, where
$w=2\pi/T>0$ is the frequency function.
A straightforward computation shows that the generator of this $U(1)-$action is given by the vector field
$$
\Upsilon =\frac{1}{w}X{_{H_0}}\in\mathcal{X}(M)\,.
$$
Now, for any function $f\in\mathcal{C}^\infty(M)$, its $U(1)-$averaging is defined in terms
of the pullback by the flow:
$$
\left\langle f\right\rangle =\frac{1}{T}\int^T_0 (\mathrm{Fl}^t_\Upsilon )^*f\,\mathrm{d}t\,.
$$
Also, an $\mathcal{S}$ operator, mapping $\mathcal{C}^\infty(M)$ into itself, is defined as
$$
\mathcal{S}(f)=\frac{1}{T}\int^T_0 (t-\pi )(\mathrm{Fl}^t_\Upsilon )^*f\,\mathrm{d}t\,.
$$
The solution to the homological equations can be expressed in terms of these operators
(see \cite{AVV13}). In
particular, the lowest order expressions for the normal forms of the perturbed Hamiltonian are
\begin{equation}\label{N1}
N_1=\left\langle H_1\right\rangle =\frac{1}{T}\int^T_0 (\mathrm{Fl}^t_{\Upsilon})^*H_1\,
\mathrm{d}t\,,
\end{equation}
and
\begin{equation}\label{N2}
N_2 =\frac{1}{2}\left\langle
\left\lbrace \mathcal{S}\left( \frac{H_1}{w} \right),H_1\right\rbrace
\right\rangle\,.
\end{equation}

\section{Invariants of the Hamiltonian flow of the harmonic oscillator with two degrees of freedom.}\label{sec-Hamflow}
Consider the harmonic oscillator with two degree of freedom on $T^\ast\mathbb{R}^2$ with coordinates $(q_1,p_1,q_2,p_2)$ and the Poisson bracket induced by the usual canonical
symplectic structure, whose Hamiltonian is
\begin{equation}\label{hamiltonian0}
H_0(q_1,p_1,q_2,p_2)=\frac{1}{2}(p^2_1+\omega^2_1q^2_1+p^2_2+\omega^2_2q^2_2)\,.
\end{equation}
The associated Hamiltonian vector field is readily found to be
$$
X_{H_0} = p_1\frac{\partial}{\partial q_1}-\omega^2_1q_1\frac{\partial}{\partial p_1}
+p_2\frac{\partial}{\partial q_2}-\omega^2_2q_2\frac{\partial}{\partial p_2}\,.
$$
The integral curves of $X_{H_0}$, $c:I\subset\mathbb{R}\to T^\ast\mathbb{R}^2$, can be parametrized as $c(t)=(q_1(t),p_1(t),q_2(t),p_2(t))$, and satisfy the decoupled system
(where the dots denote time derivatives)
$$
\begin{cases}
\ddot{q_1}+\omega^2_1q_1=0\\[5pt]
\ddot{q_2}+\omega^2_2q_2=0\,.
\end{cases}
$$
Hence, we have an action on $T^\ast\mathbb{R}^2\simeq \mathbb{R}^4$ given
by the (linear) flow of $X_{H_0}$:
\begin{align*}
\mathrm{Fl}^t_{X_{H_0}}
\begin{pmatrix} 
q_1\\[4pt]
p_1\\[4pt]
q_2\\[4pt]
p_2
\end{pmatrix}=&
\begin{pmatrix}
q_1\cos \omega_1t+\frac{p_1}{\omega_1}\sin \omega_1t\\[4pt]
-\omega_1q_1\sin \omega_1t+p_1\cos \omega_1t\\[4pt]
q_2\cos \omega_2t+\frac{p_2}{\omega_2}\sin \omega_2t\\[4pt]
-\omega_2q_2\sin \omega_2t+p_2\cos \omega_2t
\end{pmatrix}\,.
\end{align*}
This flow is periodic whenever $\omega_1$ and $\omega_2$ are commensurable so, by a suitable 
rescaling in time, we actually have a $U(1)$ action. In particular, if 
$\omega_1,\omega_2\in\mathbb{Z}$ are coprime, as in the case of the $1:2$ resonance 
that we will consider (that is, $\omega_1=1$, $\omega_2=2$), then 
$\mathrm{Fl}^t_{X_{H_0}}$ is already $2\pi-$periodic. Notice that periodic orbits will be 
invariant sets under the action of this flow, so we expect to be able of finding
them by studying the invariant functions. In fact, we will restrict our attention to the set of 
invariant polynomials under the action of this flow; the reason is that any other invariant will
be a smooth function of these, as we will see below. 

It is well know that the  algebra of invariant polynomials (under the action of the
Hamiltonian flow of $X_{H_0}$) is finitely generated, see for example \cite{CKR83,CB97}. 
Moreover, the generators can be chosen as the so-called the \emph{Hopf variables}:
\begin{align*}
\rho_1 =& z_1\overline{z}_1=\omega^2_1q^2_1+p^2_1 \\
\rho_2 =& z_2\overline{z}_2=\omega^2_2q^2_2+p^2_2 \\
\rho_3 =& \mathrm{Re}\left( z^{\omega_2}_1\overline{z}^{\omega_1}_2 \right)
	   = \mathrm{Re}\left( (p_1+i\omega_1q_1)^{\omega_2}(p_2-i\omega_2q_2)^{\omega_1} \right) \\
\rho_4 =& \mathrm{Im}\left( z^{\omega_2}_1\overline{z}^{\omega_1}_2 \right) 
	   = \mathrm{Im}\left( (p_1+i\omega_1q_1)^{\omega_2}(p_2-i\omega_2q_2)^{\omega_1} \right) \,.
\end{align*}

For instance, in the case of the $1:2$ resonance we get
\begin{align}\begin{split}\label{rhos}
\rho_1 =& q^2_1+p^2_1 \\
\rho_2 =& 4 q^2_2+p^2_2 \\
\rho_3 =& p_2(p^2_1-q^2_1)+4p_1q_1q_2 \\
\rho_4 =& 2q_2(p^2_1-q^2_1)-2q_1p_1p_2\,.
\end{split}\end{align}

There exists a certain algebraic relation satisfied by the $\rho$ variables, namely:
$$
\rho^2_3 + \rho^2_4=\rho^{\omega_2}_1\rho^{\omega_1}_2\,,\quad \rho_1,\rho_2\geq 0\,,
$$
which is the equation of a singular algebraic surface in $\mathbb{R}^4$. For the
particular case of the $1:2$ resonance, this is
\begin{equation}\label{rhoeqsbis}
\rho^2_3 + \rho^2_4=\rho^{2}_1\rho_2\,,\quad \rho_1,\rho_2\geq 0\,.
\end{equation}

Since (by a suitable rescaling) the action on $T^\ast \mathbb{R}^2\simeq \mathbb{R}^4$
of the flow of $X_{H_0}$ can be seen as a smooth $U(1)-$action, the group $U(1)$ is compact,
and the orbit space $\mathbb{R}^4/U(1)$ only contains finitely many orbit types (we will
consider the geometric structure of this orbit space later on), we can apply
the result in \cite{Sch75}, which tells us that the \emph{smooth} observables invariant under
the action of $U(1)$ are \emph{smooth} functions of the polynomial generators 
$(\rho_1,\rho_2,\rho_3,\rho_4)$.

\section{Second-order normal form of the Hamiltonian}\label{sec-normal}

In order to prove analytically the existence of periodic orbits for the Wilberforce pendulum and determine their stability, we compute the second order normal form in the case of a quartic interaction and $1:2$ resonance:
\begin{align}\label{wilber}
H(q_1,p_1,q_2,p_2)=& H_0+ \varepsilon H_1 \\
=& \frac{1}{2}(p^2_1+q^2_1+p^2_2+4q^2_2) + \nonumber
\varepsilon q_1^2q_2^2\,,
\end{align}

Since $\{H_0,N_i\}=\mathcal{L}_{X_{H_0}}N_i=0$, the first and second order normal forms
are invariant under the $U(1)-$action induced by the flow of $H_0$; we will take the
quotient of the phase space by this action and get the corresponding Hamiltonian on the 
reduced phase space in the next section. An important feature of this reduction process 
is that this reduced Hamiltonian will be a function of only three among the invariant 
generators $(\rho_1,\rho_2,\rho_3,\rho_4)$. Previous to reduction, we compute in this section
the expressions of $N_1$ and $N_2$.

Notice that the Hamiltonian flow $\mathrm{Fl}^t_{X_{H_0}}$ in this case
is given by
\begin{align}\label{hamflow}
\mathrm{Fl}^t_{X_{H_0}}
\begin{pmatrix} 
q_1\\[4pt]
p_1\\[4pt]
q_2\\[4pt]
p_2
\end{pmatrix}=&
\begin{pmatrix}
q_1\cos t+p_1\sin t\\[4pt]
-q_1\sin t+p_1\cos t\\[4pt]
q_2\cos 2t+\frac{p_2}{2}\sin 2t\\[4pt]
-2q_2\sin 2t+p_2\cos 2t
\end{pmatrix}\,
\end{align}
and it is $2\pi-$periodic. The second-order normal form of the Wilberforce oscillator is
$\displaystyle H_0 + \varepsilon N_1 +\frac{\varepsilon^2}{2} N_2$, where $N_1$ and $N_2$
are given by \eqref{N1}, \eqref{N2}. The computations are straightforward but tedious, and are best done using a computer algebra system (CAS). We have found the CAS Maxima very useful in this regard, and we have written a small Maxima package for this kind of computations, called
\texttt{pdynamics}, which is available at \url{https://github.com/josanvallejo/pdynamics}.

The resulting normal form sub-Hamiltonians, already written in the Hopf variables, are as
follows:
$$ 
H_0=\frac{1}{2}(\rho_1+\rho_2)\,,
$$
for the unperturbed part, and
\begin{equation}\label{n1}
N_1=  \langle H_1 \rangle =
\frac{1}{16}\rho_1\rho_2
\end{equation}
and 
\begin{equation}\label{n2}
N_2=  \left\langle \left\{  S\left( H_1\right),H_1\right\}\right\rangle = -\frac{1}{768}(5\rho_1\rho_2^2+4\rho_3^2+16\rho_4^2)\,,
\end{equation}
for the first and second-order perturbations, respectively.

We will make use of these explicit expressions in the following sections, to determine the
existence of stable periodic orbits in the dynamics of the Wilberforce oscillator.

\section{Constructing the reduced phase space}\label{stability}

\begin{figure}[h]
\centering
\includegraphics[scale=0.5]{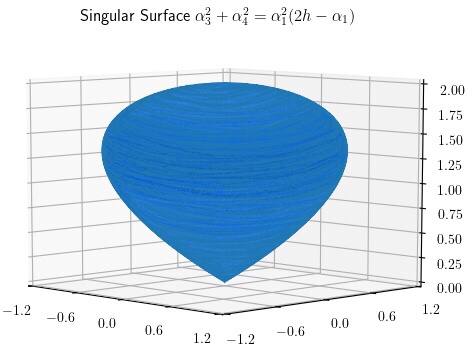} 
\caption{Reduced phase space of the 1:2 resonance.}
\label{fig3}
\end{figure}

We begin by identifying the geometry of the the reduced phase space. 
Then, we find an explicit expression for the
reduced Hamiltonian, that is, the normal form Hamiltonian 
$N=H_0 +\varepsilon N_1 + \frac{1}{2}\varepsilon^2 N_2+ O(\varepsilon^3)$ 
restricted to the reduced phase space. We follow the technique described in \cite{CKR83,Cus94} to 
prove that \eqref{rhoeqsbis} and the condition of constant energy $H_0=h >0$, give the algebraic 
description of the reduced phase space. We use a result in \cite{Poe76}, 
which states that the basic invariant 
polynomials separate the orbits of the Hamiltonian flow 
$\mathrm{Fl}^t_{X_{H_0}}$. In our case this implies\footnote{Here we collectively denote
$(q_1,p_1,q_2,p_2)$ by $(q,p)$.} that the equality 
$(\rho_1(q,p),\ldots,\rho_4(q,p))=(\rho_1(q',p'),\ldots,\rho_4(q',p'))$ holds if and only if 
$(q,p)$ and $(q',p')$ belong to the same orbit. Thus, it is enough to prove that for every 
$(u_1,u_2,u_3,u_4)$ such that $u^2_3+u^2_4= u^{2}_1u_2$, its inverse image under the map
$(q,p)\mapsto (\rho_1(q,p),\ldots,\rho_4(q,p))$ is precisely a single orbit of the flow
$\mathrm{Fl}^t_{X_{H_0}}$. For instance, if $u_2=0$ then $\rho_2(q,p)=0$ and necessarily
$q_2=0=p_2$ (from \eqref{rhos}). This, in turn, implies that $\rho_3=0=\rho_4$ so
we have the inverse image of $(u_1,0,0,0)$, where $u_1\geq 0$, which is the set
$\{(q_1,p_1,0,0)\in\mathbb{R}^4: q^2_1+p^2_1=u_1\}$, and this is clearly an orbit of
$\mathrm{Fl}^t_{X_{H_0}}$. The remaining cases can be done along similar lines, and
will not be repeated here. The reduced phase space is then given by the set of equations
$$
\begin{cases}
\rho^2_3 + \rho^2_4=\rho^2_1\rho_2\,,\quad \rho_1,\rho_2\geq 0 \,,\\
\rho_1+\rho_2=2h\,,
\end{cases}
$$
that is,
\begin{equation}\label{algsur}
\rho^2_3 + \rho^2_4=\rho^2_1(2h-\rho_1)\,,\quad 0\leq\rho_1\leq 2h\,.
\end{equation}
As mentioned above, \eqref{algsur} is the equation of a singular algebraic surface 
$S\in\mathbb{R}^3$. Topologically, this surface is a pinched sphere with a singularity
at the point $(\rho_1,\rho_3,\rho_4)= (0, 0,0)$ (see Figure \ref{fig3}).

One of the most important results in the theory is a theorem by Moser (see \cite{Mos70,CKR83}),
which can be stated as follows:
Let $H=H_0+\varepsilon H_1$ be a perturbed Hamiltonian, with $S$ the hypersurface
$H_0=h$. Suppose that the orbits of the Hamiltonian flow $\mathrm{Fl}^t_{X_{H_0}}$ are all periodic 
with period $2\pi$ and let $M_h$ be the quotient with respect to the
induced $U(1)-$action on $S$. Then, to every \emph{non-degenerate} critical point
$p\in M_h$ of the restricted averaged perturbation
$\left. N_1\right|_S=\left. \left\langle H_1\right\rangle\right|_{M_h}$
corresponds a \emph{periodic} trajectory of the full Hamiltonian vector field $X_H$, that branches
off from the orbit represented by $p$ and has period close to $2\pi$.

In order to apply this result, we must first characterize the critical points of Hamiltonian vector fields in the the reduced space. First, observe that the commutator relations among generators $(\rho_1,\rho_2, \rho_3, \rho_4 )$ are given by
\begin{eqnarray}\label{relconm}
  \{\rho_1,\rho_2 \} &=& 0, \ \ \ \ \{\rho_1,\rho_3 \} = -4\rho_4, \ \ \ \ \{\rho_1,\rho_4 \} = 4\rho_3, \nonumber \\
  \{\rho_2,\rho_3 \} &=& 4\rho_4, \ \ \ \ \{\rho_2,\rho_4 \} = -4\rho_3, \nonumber\\
  \{\rho_3,\rho_4 \} &=&-4 \rho_1(\rho_1-2\rho_2).
\end{eqnarray}
Renaming the variables $\rho_3=x$, $\rho_4=y$, and $\rho_1=z$, these relations induce a Poisson bracket on the three dimensional Euclidean space $\mathbb{R}^3=\{(x,y,z) \}$ given by
\begin{equation}\label{poibrac}
\{f, g\} = 2\left\langle\nabla g,\nabla f\times\nabla F\right\rangle\,, 
\end{equation}
where $F$ is the function
\begin{equation}\label{funcbrac}
F(x,y,z)=x^2+y^2-z^{2}(2h-z)\,,
\end{equation}
and the symbols $\left\langle\cdot ,\cdot\right\rangle$, $\times$, $\nabla$ stand for the usual inner product, cross product and nabla operator in $\mathbb{R}^3$, respectively. 
Hence, for any $f\in C^\infty(\mathbb{R}^3)$, its Hamiltonian vector field is given by
\begin{equation}\label{Hamvect}
X_f= 2 \nabla f \times \nabla F\,.
\end{equation}
It follows directly from definition \eqref{poibrac}  that the function $F(x,y,z)$ \eqref{funcbrac} 
is a Casimir of the Poisson structure \eqref{poibrac}. Thus, the symplectic leaves of the 
corresponding foliation are precisely the connected components of level sets of $F$.
If we define the mapping $P:\mathbb{R}^4 \to \mathbb{R}^4$ by
$$
P(\rho_1,\rho_2,\rho_3,\rho_4) = (\rho_3,\rho_4,\rho_1)\,,
$$
we get that $P$ is a Poisson map and $P(H_0^{-1}(h))= F^{-1}(0).$ Moreover,
\begin{equation*}
\left(P\circ \mathrm{Fl}^t_{X_{H_0}}\right)
\begin{pmatrix}
q_1\\
p_1\\
q_2\\
p_2
\end{pmatrix}
= P\begin{pmatrix}
\rho_1(p_1,q_1,p_2,q_2)\\
\rho_2(p_1,q_1,p_2,q_2)\\
\rho_3(p_1,q_1,p_2,q_2)\\
\rho_4(p_1,q_1,p_2,q_2)
\end{pmatrix}\,.
\end{equation*}
Therefore, the reduced space is contained in a symplectic leaf of 
$F^{-1}(0)\subset \mathbb{R}^3$ . Let us denote by $M_h$ the reduced space. Then, a realization
of it as a smooth manifold\footnote{Notice that the condition $z>0$ removes the singularity at
the origin.} is given by
\begin{equation}
M_h= F^{-1}(0) \text{ and } 0 <  z \leq 2h \,.
\end{equation}
Any function $f\in C^\infty(\mathbb{R}^3)$ defines a Hamiltonian vector field $\widetilde{X}_f$ 
on $M_h$ by the restriction of \eqref{Hamvect}:
\begin{equation*}
\widetilde{X}_f:= \left. (2\nabla f\times\nabla F)\right|_{M_h}.
\end{equation*}
It also follows from \eqref{Hamvect} that the Hamiltonian vector field $\widetilde{X}_f$ has a critical point at the point $p\in M_h$ if and only if either $\nabla f(p)$ is orthogonal at $p$
to the reduced space $M_h$, or $\nabla f(p)=0$.

\begin{figure}[h]
\centering
\includegraphics[scale=0.5]{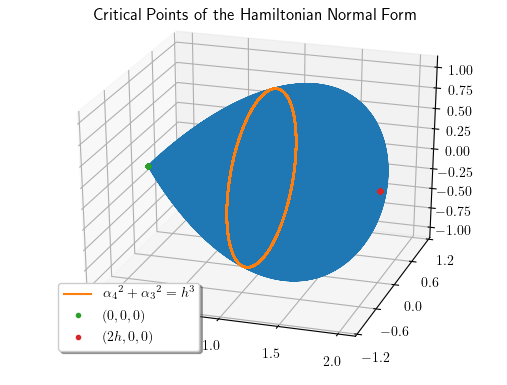}
\caption{Critical points of the first-order normal Hamiltonian $N_1$.}
\label{fig4} 
\end{figure}

Next, we describe how to obtain the reduced Hamiltonian vector field corresponding to a function $G 
\in C^\infty(\mathbb{R}^4)$ such that $\{H_0, G\}=0$. As discussed above,  
$G$ can be expressed in terms of the Hopf variables: $G = G(\rho_1,\rho_2,\rho_3,\rho_4)$.
Writing  $\rho_1=z, \rho_2=2h-z, \rho_3=x $ and $\rho_4=y$, we obtain the function 
$Q(x,y,z)= G(z,2h-z,x,y)$. Thus, the reduced Hamiltonian vector field associated to $G$ is the 
vector field
$$
\widetilde{X}_G=\left.(  2 \nabla Q \times \nabla F )\right|_{M_h}\,.
$$
This expression allows us to compute the critical points of the reduced vector field associated to 
the first-order normal form $N_1(\rho_1,\rho_2,\rho_3,\rho_4)$ given in \eqref{n1}.
Letting as above $K(x,y,z)=N_1(z,2h-z,x,y)$, we get
\begin{eqnarray}
K(x,y,z)&=&\frac{1}{16}(2hz-z^2)\, . \label{n1rest}
\end{eqnarray}
Hence, the reduced vector field is
\begin{equation}\label{redvector}
\widetilde{X}_{N_1}=\left.( 2 \nabla F \times \nabla K   )\right|_{M_h}.  
\end{equation}
As we pointed out above, the critical points of \eqref{redvector} are those points $p\in M_h$ such that either $\nabla K(p)=0$ or $\nabla K(p)$ is orthogonal to $M_h$ (parallel to $\nabla F (p)$). 
It is immediate to calculate
\begin{equation*}
\nabla K = \left(0,0,\frac{1}{8}(h-z) \right)\,.
\end{equation*}
It follows form here that $\nabla K(p)$ is orthogonal to $M_h$ at the point $(0,0,2h)$  and that $\nabla K(p)=0$ if $z=h$. Thus, the reduced vector field $\widetilde{X}_{N_1}$ has a critical point at $(0,0,2h)$ and a curve of critical points given by  
$\Gamma_h=\{(x,y,z)| x^2+y^2 = h^{3}\text{ and } z=h \}$, see Figure \ref{fig4} (where the singular point $(0,0,0)$ is also shown).

Consider the critical point $(0,0,2h)$. By a straightforward computation, we get 
\begin{equation*}
\frac{\partial F}{\partial z}(0,0,2h)=4h^2\neq 0.
\end{equation*}
By the implicit function theorem, $z=\psi(x,y)$ with $\psi$ a smooth function at $(0,0)$ 
satisfying $\psi(0,0)=2h$ and $F(x,y,\psi(x,y))=0$. Therefore, the function $K$ in
\eqref{n1rest} has the form $\widetilde{K}=K(\psi(x,y))$ in a neighborhood of $(0,0,2h)$. 
Another immediate computation shows that
\begin{equation*}
\mathrm{Hess}(\widetilde{K}(0,0)) =\frac{1}{16h^2}>0\,.
\end{equation*}
Thus, the critical point $(0,0,2h)$ is non-degenerate and Moser's theorem (see the version 
presented as Theorem 6.4 in \cite{CKR83}) implies that, for small enough $\varepsilon$, the 
Wilberforce oscillator has a unique stable periodic orbit $\gamma_\varepsilon$ with energy $h$ 
through each point $p(\varepsilon)$,  sufficiently close to $(0,0,2h)$, with period 
$T(\varepsilon)$, such that $H_0(p(\varepsilon))\to h$ and $T(\varepsilon) \to 2\pi$.

\section{Stability analysis at the degenerate points}

We can not apply Moser's theorem to the curve of critical points of the preceding section,
$\Gamma_h$, because they are degenerate (non-isolated). 
In order to determine if some periodic orbits arise from
some points of $\Gamma_h$ we must resort to the second-order normal form of the Hamiltonian
\eqref{wilber}, which will be regarded as a perturbed Hamiltonian $H_0+\varepsilon (N_1+\varepsilon N_2)$ on its own.

Thus, we consider the $\varepsilon-$dependent function $K_\varepsilon(x,y,z) = \left.(N_1 +\varepsilon N_2)\right|_{M_h}$. By arguments similar to those used in the case of Moser's theorem, 
we readily see that a given point of $\Gamma_h$ generates a periodic orbit of the Wilberforce pendulum if it is a non-degenerate critical point of the Hamiltonian vector field $\displaystyle X_{ K_\varepsilon}=2 \nabla K_\varepsilon \times \nabla F $ \emph{for all} $\varepsilon$. 
Therefore, we need to look for points $p\in \Gamma_h$ such that, either $\nabla K_\varepsilon(p)=0$, or $\nabla K_\varepsilon(p)$ is parallel to $\nabla F(p)$ for all 
$\varepsilon$. A straightforward computation gives
\begin{equation*}
\nabla K_\varepsilon =\left(-\frac{\varepsilon x}{96},-\frac{\varepsilon y}{24},\frac{h-z}{8}-\frac{5\varepsilon}{768}(4h^2-8hz+3z^2) \right).
\end{equation*}

For this vector to vanish, its third component must be zero independently of $\varepsilon$,
that is, both the independent term and the coefficient of $\varepsilon$ must vanish separately.
These conditions would lead to $z=h$ and $h=0$, so the only possibility is the point $(0,0,0)$
on the particular surface $M_0$, which is the case of the singular point that we consider
in the next section.
It follows from here that $\nabla K_\varepsilon$ never vanishes for $\varepsilon \neq 0$ and
$h\neq 0$, and it is easy to check that it is parallel to $\nabla F$ only at the points 
$(0,0,\frac{2}{3}h)$ and $(0,0,2h)$, which do not belong to the curve $\Gamma_h$.
Consequently, we conclude that no points of $\Gamma_h$ (aside from the singular point $(0,0,0)$
in the case $h=0$) can generate a periodic orbit.

\section{Stability analysis at the singular point}
Recall that, in order to impose a smooth structure on the reduced space, we left aside the 
singular point $(0,0,0)$. To complete our analysis, in this section we deal with that 
particular case (which corresponds in the literature to the so-called \emph{normal mode},
$\gamma$). The existence of closed orbits will be proved by finding fixed points on a 
suitable Poincar\'e section.

Let $f_2(p_1,q_1,p_2,q_2) = \frac{1}{2}(p_2^2+4q^2_2 )$. The Hamiltonian vector field
with respect to the canonical symplectic structure on $\mathbb{R}^4$, $X_{f_2}$, has 
periodic flow with periodic $T=\pi$. This flow generates a free and proper
$U(1)-$action on $(\mathbb{R}^2-(0,0))\times\mathbb{R}^2$. For every fixed $h>0$, the 
level set $f_1^{-1}(h)$ is foliated by periodic orbits of $X_{f_2}$ and a the reduced
space is given by $M_h=f_2^{-1}(h)/U(1)$. Let us make the following change of variables
from $(p_1,q_1,p_2,q_2)$ to $(p_1,q_1,L,\theta )$:
$$
\Psi(p_1,q_1,L,\theta )=(p_1,q_1,-\sqrt{4L}\sin \theta,\sqrt{L}\cos  \theta)\,,
$$
with $L>0\,, 0<\theta<2\pi/\omega_2$.
In these coordinates, the canonical symplectic form on the domain 
$\mathbb{R}^2\times(\mathbb{R}^2-(0,0))$, given by 
$\mathrm{d}p_1\wedge\mathrm{d}q_1+\mathrm{d}p_2\wedge\mathrm{d}q_2$, 
becomes $\mathrm{d}p_1\wedge\mathrm{d}q_1+\mathrm{d}L\wedge\mathrm{d}\theta$, 
and the Hamiltonian of the Wilberforce oscillator is
\begin{equation} \label{Hamnew}
H(p_1,q_1,L,\theta)=\frac{1}{2}(p_1^2+q_1^2)+2L + \varepsilon Lq_1^2\cos^2\theta.
\end{equation}
Consider the restriction to the level set $\Sigma_h= \{(p_1,q_1,L,\theta)| L=h\}$. Since this 
level set is foliated by orbits of $X_{f_2}$, the Hamiltonian equations of \eqref{Hamnew} are
\begin{equation}\label{solsistnew}
\begin{cases}
\dot{\theta} =& 2 + \varepsilon q_1^2\cos^2\theta\,,\\[6pt]
\dot{p}_1 =& -q_1-2\varepsilon hq_1\cos^2\theta,\\[4pt]
\dot{q}_1 =& p_1\,.
\end{cases}
\end{equation}
We now construct the cross section 
$\sigma_0 = \{(p_1,q_1,h,\theta)\in \Sigma_h\,:\theta =0\}$, 
and fix the point $a=((p_1^0,q_1^0,h,0))$ on it. 
The integral curve of \eqref{solsistnew} through $a$ is:
\begin{equation}\label{snew}
\begin{cases}
\theta(t) =2t +
\varepsilon\int^t_0 q_1^2\cos^2\theta\,\mathrm{d}\tau\,,
\\
p_1(t) = p_1^0 \cos  t - q_1^0\sin  t-2\varepsilon h\int^t_0 q_1\cos^2\theta\, \mathrm{d}\tau\,,\\
q_1(t) = p_1^0\sin t + q_1^0\cos  t\,.
\end{cases}
\end{equation}
Let $T(a,\varepsilon)$ be the time elapsed between two consecutive intersections of $\sigma_0$. 
From equations \eqref{snew}, we get
\begin{equation*}
4\pi=2T(a,\varepsilon)+\varepsilon\int^{T(a,\varepsilon)}_0 q_1^2\cos^2\theta\,\mathrm{d}t\,,
\end{equation*}
so $T(a, \varepsilon)$ has the form 
\begin{equation}\label{time}
T(a, \varepsilon)=2\pi-\varepsilon\frac{\pi}{2} (q_1^0)^2+O(\varepsilon^2)\,. 
\end{equation}
Substituting \eqref{time} in \eqref{snew}, we obtain the following expression for the Poincar\'e
map determined by $\sigma_0$:
\begin{eqnarray*}
p_1(T(a)) &=& p_1^0 +\varepsilon\pi q_1 \left(\frac{1}{2}(q_1^0)^2-2 h\right) +O(\varepsilon^2),\\
q_1(T(a)) &=&q_1^0 +\varepsilon p_1 \frac{\pi}{2}(q_1^0)^2+O(\varepsilon^2).
\end{eqnarray*} 
In order to prove that there exists periodic orbits for the Wilberforce oscillator in $\Sigma_h$, 
we must show that, for each $\varepsilon$ small enough, there exist $p_1^0(\varepsilon)$ and 
$q_1^0(\varepsilon)$ such that we get a fixed point:
\begin{eqnarray*}
p_1(T(p_1^0(\varepsilon),q_1^0(\varepsilon),-h,0,\varepsilon))= p_1^0(\varepsilon), \\
q_1(T(p_1^0(\varepsilon),q_1^0(\varepsilon),-h,0,\varepsilon))= q_1^0(\varepsilon).
\end{eqnarray*}
To this end, we define the following function $F: \mathbb{R}^3\rightarrow \mathbb{R}^2$,
\begin{equation*}
F\begin{pmatrix}
p_1 \\
q_1 \\
\varepsilon
\end{pmatrix}=
\begin{pmatrix}
\pi q_1 \left(\frac{1}{2}(q_1^0)^2-2 h\right)  +O(\varepsilon) \\
p_1 \frac{\pi}{2}(q_1^0)^2+O(\varepsilon)
\end{pmatrix}\,.
\end{equation*}
First, we note that $F(0,2\sqrt{h},0)=0. $ A straightforward computation shows that
\begin{equation*}
\det \left( \left. \frac{\partial F}{\partial p_1\partial q_1} \right|_{(0,2\sqrt{h},0)} \right) = \det 
\begin{pmatrix}
\pi h & 0\\
0 & \pi h
\end{pmatrix} >0\,.
\end{equation*}
By the implicit function theorem, there exists $\delta>0$, an open neighborhood $U$ of $(0,2\sqrt{h})$, and a function $g: (-\delta,\delta)\rightarrow U$, $g(\varepsilon)=(p_1(\varepsilon),q_1(\varepsilon) )$, such that $g(0)=(0,0)$ and $F(g(\varepsilon),\varepsilon)=0$. Therefore,
\begin{eqnarray*}
p_1(T(g(\varepsilon),-h,0,\varepsilon))= p_1(\varepsilon), \\
q_1(T(g(\varepsilon),-h,0,\varepsilon))= q_1(\varepsilon).
\end{eqnarray*}
This fact proves that for each sufficiently small $\varepsilon$, the Wilberforce oscillator has a 
unique stable periodic orbit $\gamma_\varepsilon$, with energy $h$, which branches off from the 
normal mode $\gamma$.


\textbf{Acknowledgements}: MAC was partially supported by a Mexican CONACyT Research Project code CB-258302, ATM was supported by a Mexican CONACyT graduate student grant, and JAV was partially supported by a Mexican CONACyT Research Project code A1-S-19428.

\Addresses


\begin{thebibliography}{99}
\bibitem[Avenda\~no-Camacho et al(2013)]{AVV13}
Avenda\~no-Camacho~M., Vallejo~JA and Vorobjev~Yu (2013)
\textit{A simple global representation for second-order normal forms of Hamiltonian systems relative to periodic flows},
J. Phys. A: Math. Theor. \textbf{46} 395201.

 \bibitem[Avenda\~no-Camacho et al(2017)]{AVV17}
Avenda\~no-Camacho~M., Vallejo~JA and Vorobjev~Yu (2017)
\textit{A perturbation theory approach to the stability of the Pais-Uhlenbeck oscillator},
J. of Math. Phys. \textbf{58} 093501.

\bibitem[Berg and Marshall(1991)]{BM91}
Berg~RE and Marshall~TS (1991) 
\textit{Wilberforce pendulum oscillations and normal modes},
Am. J. Phys. \textbf{59} (1) 32-38.

\bibitem[de Bustos et al(2016)]{BLM16}
de Bustos~MT, L\'opez~MA and Mart\'inez~R (2016)
\textit{On the periodic orbits of the perturbed Wilberforce pendulum},
J. of Vibrations and Control \textbf{22} (4) 932-939.

\bibitem[Churchill et al(1983)]{CKR83}
Churchill~RC, Kummer~M and Rod~DL (1983)
\textit{On averaging, reduction, and symmetry in Hamiltonian systems},
J. Differ. Eq. \textbf{49} 359-414.

\bibitem[Cushman(1994)]{Cus94}
Cushman~RH (1994)
\textit{Geometry of perturbation theory}, in `Deterministic Chaos in General Relativity', edited by Hobill~D et al. Springer Verlag, 89-101.

\bibitem[Cushman and Bates(1997)]{CB97}
Cushman~RH and Bates~LM (1997)
\textit{Global Aspects of Classical Integrable Systems},
Birkh\"auser Basel.

\bibitem[Deprit(1969)]{Dep69}
Deprit~A (1969)
\textit{Canonical transformation depending on a small parameter},
Celest. Mech. \textbf{1} (1) 1-30.

\bibitem[Holmes(2007)]{Hol07}
Holmes~M (2007)
\textit{Introduction to Numerical Methods in Differential Equations},
Springer Verlag.

\bibitem[K\"opf(1990)]{Ko90}
K\"opf~U (1990) 
\textit{Wilberfore pendulum revisited},
Am. J. Phys. \textbf{58} (9) 833-839.

\bibitem[Moser(1970)]{Mos70}
Moser~J (1970)
\textit{Regularization of Kepler's problem and the averaging method on a manifold},
Comm. in Pure and Appl. Math. \textsc{XXIII} 609-636.

\bibitem[Ott(2002)]{Ott02}
Ott~E (2002)
\textit{Chaos in dynamical systems},
2nd~edn. Cambridge UP.

\bibitem[Plav\v{c}i\'c et al(2009)]{PZB09}
Plav\v{c}i\'c~M, \v{Z}upanovi\'c~P and Bona\v{c}i\'c-Lo\v{s}i\'c~Z (2009) 
\textit{The resonance of the Wilberforce pendulum and the period of beats},
Lat. Am. J. Phys. Educ. \textbf{3} (3) 547-550.

\bibitem[Po\`enaru(1976)]{Poe76}
Po\`enaru~V (1076)
\textit{Singularit\'es $C^\infty$ en pr\'esence de sym\'etrie},
Lecture Notes in Mathematics \textbf{510}, Springer Verlag.

\bibitem[Schwarz(1975)]{Sch75} 
Schwarz~G (1975)
\textit{Smooth funtions invariant under the action of a compact Lie group},
Topology \textbf{14} 63-68.

\bibitem[Strogatz(2015)]{Str15}
Strogatz~SH (2015)
\textit{Nonlinear dynamics and chaos},
2nd~edn. Westview Press.


\end{thebibliography}
\end{document}